\documentclass[aps,twocolumn,showpacs,superscriptaddress]{revtex4}
\usepackage[dvips]{graphicx}
\usepackage{amsmath}
\usepackage{amssymb}
\usepackage{dcolumn}
\begin{document}
\title{Velocity distributions in dilute granular systems}
\author{J.\ S.\ van Zon}
\email{jvzon@nat.vu.nl}
\author{F.\ C.\ MacKintosh}
\email{fcm@nat.vu.nl} \affiliation{Division of Physics and
Astronomy, Vrije Universiteit, 1081 HV Amsterdam, The Netherlands}
\affiliation{Kavli Institute for Theoretical Physics, University of
California, Santa Barbara, CA 93106}
\date{\today}
\begin{abstract}
Motivated by recent experiments reporting non-Gaussian velocity distributions in driven dilute granular materials, we study by numerical simulation the properties of inelastic gases as functions of the coefficient of restitution $\eta$ and concentration $\phi$ with various heating mechanisms. We show that there are marked, qualitative differences in the behavior for uniform heating (as is frequently assumed theoretically) and for particle systems driven at the boundaries of the container (as is frequently done in experiments). In general, we find Gaussian velocity distributions for uniform heating and non-Gaussian velocity distributions for boundary heating. Furthermore, we demonstrate that the form of the observed velocity distribution is governed primarily by the coefficient of restitution $\eta$ and $q=N_H/N_C$, the ratio between the average number of heatings and the average number of collisions in the gas. The differences in distributions we find between uniform and boundary heating can then be understood as different limits of $q$, for $q \gg 1$ and $q \lesssim 1$ respectively. Moreover, we demonstrate that very similar behavior is found for a simple model of a gas of inelastic particles with no spatial degrees of freedom.
\end{abstract}
 \maketitle

\section{Introduction}

Granular materials consisting of macroscopic particles or grains can exhibit behavior reminiscent of conventional phases of matter. Sand, for instance, can flow like a liquid under some conditions. Dilute granular systems, or {\em gases}, have been extensively studied both experimentally and theoretically, in large part as simple model systems exhibiting nonequilibrium and dissipative behavior. These systems are intrinsically dissipative and out of equilibrium, even though it is tempting to apply such equilibrium notions as temperature. Since the collisions in such a gas are inelastic, it is necessary to supply energy, or to drive them in order to maintain a gas-like steady state. Otherwise, {\em inelastic collapse} can occur, in which all motion ceases after only a finite time\cite{mcnamara, kadanoff1}. In principle, it is possible to heat or drive the system uniformly throughout the container ({\em uniform heating}), as has been done in simulations\cite{mackintosh, moon}, and as is assumed in analytic theories\cite{ernst}. In experiments, however, one usually drives a granular gas by shaking or vibrating the walls of the container. Such {\em boundary heating} means that the energy is inserted in a spatially inhomogeneous way\cite{luding, losert, kadanoff2,rouyer, kudrolli}. As a consequence, the gas will develop a gradient in density and mean kinetic energy.

Even for uniform heating, however, significant deviations from equilibrium gases, e.g., in density correlations, are observed\cite{mackintosh}. It is often assumed that in the bulk, where the boundaries are far away and spatial gradients are small on the scale of the mean free path of the particles, both heating methods should give the same behavior. Here, we show that there are striking differences in the behavior of granular gases heated uniformly or through the boundary.

There has been much interest recently in another aspect of these non-equilibrium gases: namely, the velocity distributions. This is in part because they deviate from the Gaussian distributions that one would expect if the collisions were elastic. It has been suggested that the velocity distributions can be described by a sort-of {\em stretched} Gaussian, of the form $P(v)=C \exp [-\beta(v/\sigma)^{\alpha}]$, where $\sigma=\langle v^2\rangle ^{\frac{1}{2}}$ is sometimes called the granular temperature. Rouyer and Menon \cite{rouyer} have reported such a distribution with exponent $\alpha=1.5$ over the whole observed range of velocities, which was unaffected by changes in amplitude and frequency of driving. This observation was particularly intriguing, given the theoretical predictions obtained before by Van Noije and Ernst \cite{ernst}. They developed a kinetic theory that predicts a high-velocity tail described by a distribution with an exponent $\alpha=\frac{3}{2}$.

Unfortunately, results for velocity distributions are never unambiguous. Both in simulation and experiment, different setups and driving mechanisms usually give different behavior of the velocity distribution. For a setup where particles on a horizontal plate were driven in the vertical direction, Olafsen and Urbach \cite{olafsen} found a crossover from exponential to Gaussian distributions as the amplitude of the driving was increased. The result of Rouyer and Menon \cite{rouyer} was obtained for a different configuration where particles were confined between two vertical plates and driven in the vertical direction. Although the exponent $\alpha=1.5$ they find is reminiscent of theoretical results obtained by Van Noije and Ernst \cite{ernst}, both results are actually inconsistent in that the exponent $\alpha=\frac{3}{2}$ can only describe the high-velocity tail of the distribution. Such a high-velocity tail was actually  obtained by Moon et al. \cite{moon}. Using a simulation with a variation on uniform heating they find a velocity distribution that has an exponent of $\alpha=2.0$ for low velocities, but crosses over to an exponent of $\alpha=1.5$ for the high-velocity tail. The results of Rouyer and Menon are also limited in the sense that they were obtained for almost elastic particles with $\eta=0.93$. Blair and Kudrolli \cite{kudrolli} use a different setup where particles move along an inclined plane. Friction with the plane during collisions reduces the coefficient of restitution to $\eta\approx0.5$. They find the distribution with exponent $\alpha=1.5$ only in the very dilute case. Otherwise, the distributions deviate strongly from both Gaussian and the distribution obtained by Rouyer and Menon. It is interesting to notice that for the denser case Blair and Kudrolli find a distribution with a crossover much like the type observed by Moon et al.

At present there is no agreement on what the velocity distribution of a granular gas looks like exactly nor is it clear what causes the velocity distributions to deviate from a Gaussian. Puglisi et al. \cite{puglisi} have suggested that the deviations are caused by the spatial correlations in the gas. They propose that for a region of uniform density the velocity distribution of the gas actually is Gaussian, with a density dependent width. The spatial correlations cause density fluctuations and they claim that the non-Gaussian distributions arise as an average over the velocity distributions over these regions of different density. It is true that an average over Gaussian distributions with different widths in general yields a stretched Gaussian, but experiments performed by Olafsen and Urbach \cite{olafsen} showed that this does not happen in granular gases. They find that even within small windows of uniform density the velocity distributions remained non-Gaussian. Here, we show that non-Gaussian behavior arises even in a simple model with no spatial degrees of freedom.

We study behavior of the velocity distributions of the granular gas as a function of $\phi$, the area fraction, and $\eta$, the coefficient of restitution. Specifically, we consider the effect on the velocity distribution of driving the gas by heating uniformly, as is assumed in theory and many prior simulations, and by heating through a boundary, as is done in most experiments. We will show that there exists clear qualitative difference between the velocity distributions for uniform and boundary heating. Furthermore, we will show that there is no evidence for a universal velocity distribution with a constant exponent $\alpha=1.5$. We demonstrate instead that a family of distributions showing apparent exponents covering a wide range of values $\alpha<2$ is expected, depending on both material and experimental conditions. Furthermore, we show that the velocity distribution is governed primarily by the relative importance of collisions to heating, \emph{i.e.}, the way in which energy flows through the system of particles. Specifically, we introduce a new parameter $q=N_H/N_C$, which measures the ratio between mean numbers of heating events $N_H$ and mean number of collisions $N_C$ experienced by a typical particle. These theoretical observations can explain both the observed non-Gaussian behavior, as well as the ambiguities in the experimental and theoretical literature on dissipative gases to date. We also demonstrate that the behavior of the velocity distributions can be captured quantitatively by a simple model that takes only $\eta$ and $q$ into account, with no spatial degrees of freedom.\\

\section{Numerical Simulation}
We use an event-driven algorithm to simulate $N$ particles of radius $r$ moving in a two-dimensional box. Particles gain energy by heating and lose energy through inelastic collisions. When two particles $i$ and $j$ collide their final velocities depend on their initial velocities in the following way:
\begin{equation} 
\mathbf{v}_{i}^{\prime}=\mathbf{v}_{i}-\frac{1+\eta}{2}(\mathbf{v}_{i}
\cdot \mathbf{\hat{r}}_{ij}-\mathbf{v}_{j} \cdot
\mathbf{\hat{r}}_{ij})\mathbf{\hat{r}}_{ij},
\label{eqn:inelas}
\end{equation}
where $0\leq \eta <1$ is the coefficient of restitution and $\mathbf{\hat{r}}_{ij}$ is the unit vector connecting the centers of particles $i$ and $j$.

For uniform heating we adapted an one-dimensional algorithm described in \cite{mackintosh}. When heating uniformly, each individual particle is heated by adding a random amount to the velocity of each particle during a time step $\Delta t$:
\begin{equation}
\mathbf{v}_{i}(t + \Delta t)=\mathbf{v}_{i}+\sqrt{h \Delta
t}\mathbf{f}(t),
\label{eqn:uniheat}
\end{equation}
where $\mathbf{f}(t)$ is a vector whose components are uniformly distributed between $-\frac{1}{2}$ and $\frac{1}{2}$ and $h$ is proportional to the heating rate. After heating the system is transferred to the center-of-mass frame. Particles move in a box with sides $L=2$. We use periodic boundary conditions to simulate  bulk behavior. The time step $\Delta t$ is chosen in such a way that on average the number of collisions per time step is less than one. It should be noted that this heating mechanism is significantly different from the spatially homogeneous heating used in some experiments \cite{olafsen}. In the experiment all particles feel the same forcing, so the motion of the neighboring particles is strongly correlated in space and time. In uniform heating however, all particles are independently driven by a stochastic source and as a consequence correlations are very weak.

When heating through the boundaries, particles gain velocity upon collision with the boundary. For simplicity, we assume that the collision with the boundary is elastic. In that case, a collision occurs by reflecting $\mathbf{v}_{\perp}$, the component of the velocity perpendicular to the boundary. Heating occurs by adding a random amount of velocity to $\mathbf{v}_{\perp}$. Then after
collision with the boundary one has:
\begin{equation}
\mathbf{v}_{i}^{\prime}=\mathbf{v}-2\mathbf{v}_{\perp}+\sqrt{h}
\mathbf{f}(t).
\end{equation}
Particles move in a circular box of radius $R=1$. A symmetrical container has the advantage that it allows us to examine density and granular temperature gradients along a single coordinate $r$, the distance from the center of the box, as in the one-dimensional case \cite{kadanoff2}.

We start the simulation by distributing the particles uniformly over the box. When using boundary heating, we give each particle a small, uniformly distributed velocity to enable particles to reach the boundary. Then particles are heated and we allow the system to reach a steady state before taking data. For both uniform heating and boundary heating, data is taken periodically every time step $\Delta t$. For uniform heating, data is taken when the particles are heated, so $\Delta t$ equals the time between heating events.

\section{Simulation results: clustering}
\begin{figure}[hb]
  \centering
  \includegraphics[width=\columnwidth]{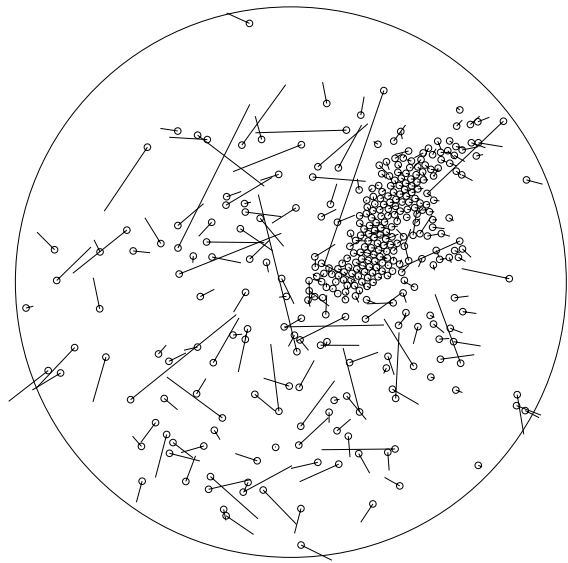}
  \caption{Snapshot of a clustered state for $N=350$,
$\phi=0.05$ and $\eta=0.6$. The circles indicate the current
positions of the particles, while the lines indicate the
direction and magnitude of their velocity.} \label{fig:cluster}
\end{figure}

\begin{figure}[ht]
  \centering
  \vspace{5mm}
  \includegraphics[width=\columnwidth]{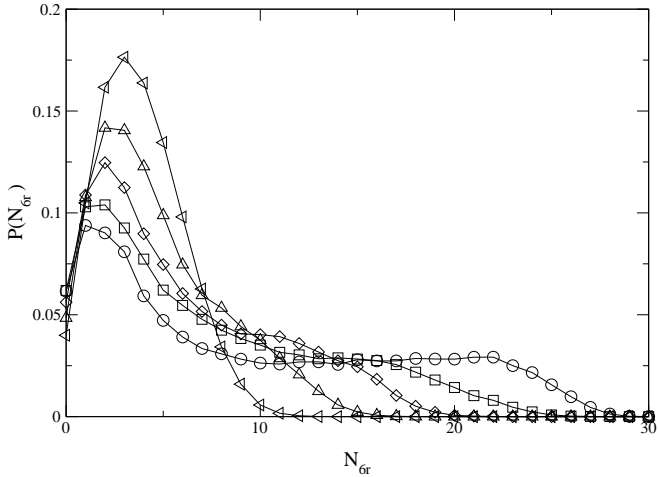}
  \caption{Number of neighbors within a distance $6r$ of a given particle
for $N=350$, $\phi=0.1$ and $\eta=0.7$($\circ$),
$0.75$($\square$), $0.8$($\diamond$), $0.85$($\triangle$) and
$0.9$($\triangleleft$). On average $N_{6r}=3.6$ for $\phi=0.1$.}
\label{fig:densityhist}
\end{figure}

One of the first striking differences between uniform heating and boundary heating is clustering. When heating through the boundary, a stable liquid-like cluster surrounded by a hot gaseous state will form for low coefficients of restitution $\eta$ or high area fraction $\phi$. This occurs as particles are compressed in the center of the box by the pressure of particles moving in from the boundary. As the cluster grows in size, it can no longer be destroyed by the impact of high velocity particles and the cluster remains stable. A typical example is shown in Fig.~\ref{fig:cluster}. We do not find these clusters when heating uniformly. This is because all particles are heated all the time, which prevents the collapse to a cluster. In experiments that find clusters in homogeneously driven systems \cite{olafsen} this is different, because here the driven motion of neighboring particles is highly correlated. The formation of a cluster spoils measurement of the velocity distribution because the system is no longer in a pure gas-like state. To avoid values of $\phi$ and $\eta$ corresponding to the formation of clusters in our simulation, we constructed a phase diagram. We did this by counting for every particle the number $N_{6r}$ of neighbors with their center within a distance $6r$ from that particle. When the gas is in a hexagonal close packed state $N_{6r}=32$. We obtained the distribution $P(N_{6r})$ for different values of $\phi$ and $\eta$. An example for $N=350$ and $\phi=0.1$ is shown in figure \ref{fig:densityhist}.


For $\eta=0.9$ the distribution corresponds to a state with the particles uniformly distributed over the box and the peak of the distribution at the mean value $\overline{N}_{6r}=3.6$. For $\eta=0.7$ the distribution becomes bimodal, with a broad peak at high $N_{6r}$ corresponding to the densely-packed cluster and a peak at $N_{6r}=1$ corresponding to the surrounding dilute gas. The distribution shows a continuous transition for $\eta$ in between, which makes it hard to pinpoint an exact value of $\eta$ for which the gas enters the clustered state. Still, by looking at the shape of the distributions, it can be argued that the transition occurs somewhere between $\eta=0.75$ and $\eta=0.85$. This was repeated for different values of $\phi$, which allowed us to determine a sort of phase diagram. Specifically, we determined the limit of a pure gas-like phase, and all results presented below were obtained in this state.

\begin{figure}[ht]
  \centering
  \vspace{0.3cm}
  \includegraphics[width=\columnwidth]{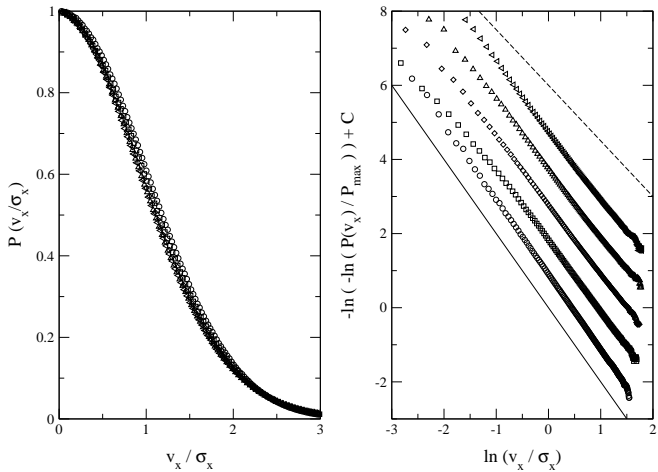}
\caption{(a) $P(v_x/\sigma_x)$. (b)
$-\ln\{-\ln[P(v_x/\sigma_x)]\}$ versus $\ln(v_x/\sigma_x)$. Data for both figures is taken for $N=350$ and for $\phi=0.02$ and $\eta=0.8$($\circ$), $0.6$($\square$), $0.4$($\diamond$), $0.2$($\triangle$), $0.1$($\triangleleft$). A Gaussian is shown as a solid line with slope $-2$ and the distribution obtained by Rouyer and Menon is shown as the dashed line with slope $-1.5$.} \label{fig:uniform_phi}
\end{figure}
\begin{figure}[hb]
    \centering
    \includegraphics[width=\columnwidth]{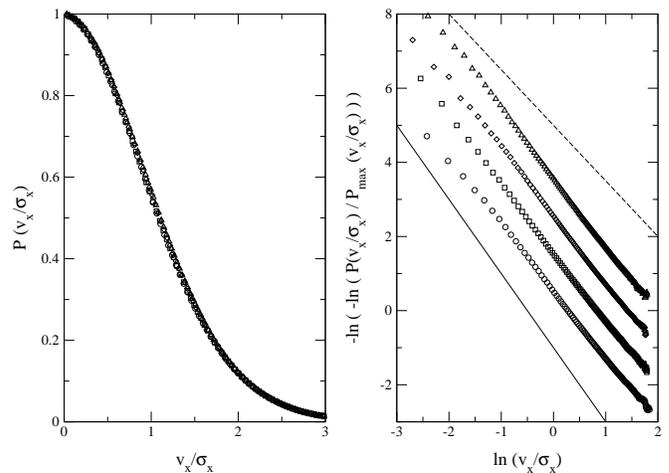}
\caption{(a) $P(v_x/\sigma_x)$. (b)
$-\ln\{-\ln[P(v_x/\sigma_x)]\}$ versus $\ln(v_x/\sigma_x)$. The
dashed lines have slope $-2$ and $-1.5$. Data for both figures is
taken for $N=350$ and for $\eta=0.2$ and $\phi=0.1$($\circ$),
$0.05$($\square$), $0.02$($\diamond$), $0.01$($\triangle$). }
\label{fig:uniform_eta}
\end{figure}
\begin{figure}[ht]
\centering \vspace{5mm}
\includegraphics[width=\columnwidth]{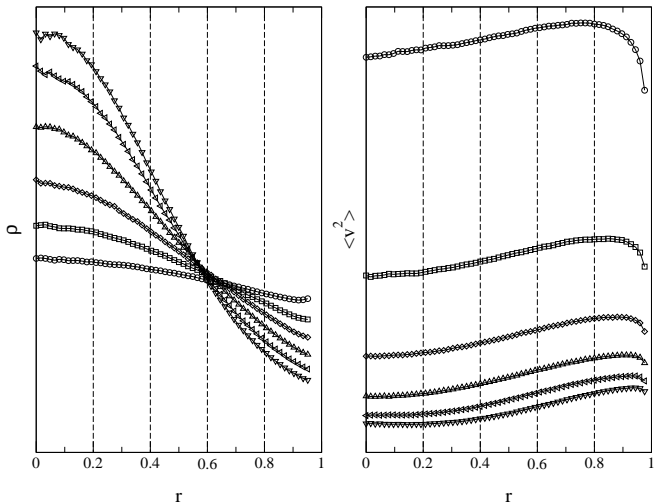}
\caption{(a) The average number density $\rho$ as a function of distance $r$ to the center of the box. Data taken for $N=350$, $\phi=0.02$ and $\eta=0.9$($\circ$), $0.8$($\square$), $0.7$($\diamond$), $0.6$($\triangle$), $0.5$($\triangleleft$) and $0.4$($\triangleright$). (b) The mean kinetic energy $\langle v^2\rangle $ per particle as a function of $r$, for the same values of $\phi$ and $\eta$. The dashed lines indicate the concentric rings, within which the velocity distributions were separately calculated.} \label{fig:profile_phi}
\end{figure}

\section{Simulation results: velocity distributions}
The velocity distributions $P(v_x)$ for uniform heating are shown in Figs.\ \ref{fig:uniform_phi} and \ref{fig:uniform_eta}. The velocity component $v_x$ is scaled by $\sigma_x=\langle v^2_x\rangle ^{\frac{1}{2}}$ and the maximum of the distribution $P(v_x/\sigma_x)$ is scaled to be unity. For a broad range of the parameters $\phi$ and $\eta$ the velocity distributions are very close to Gaussian. For $\eta=0.8$ the velocity distributions can be fitted by a distribution with $\alpha=2.0$. This decreases only slightly for $\eta=0.1$, which can be fitted by a distribution with $\alpha=1.9$. Values of $\alpha$ are found to be independent of $\phi$. These exponents are constant over the entire observed range of velocities and we see no high-velocity tails with $\alpha=1.5$ for the range of $\phi$ and $\eta$ we examined.

\begin{figure}[hb]
\centering \vspace{0.5cm}
\includegraphics[width=\columnwidth]{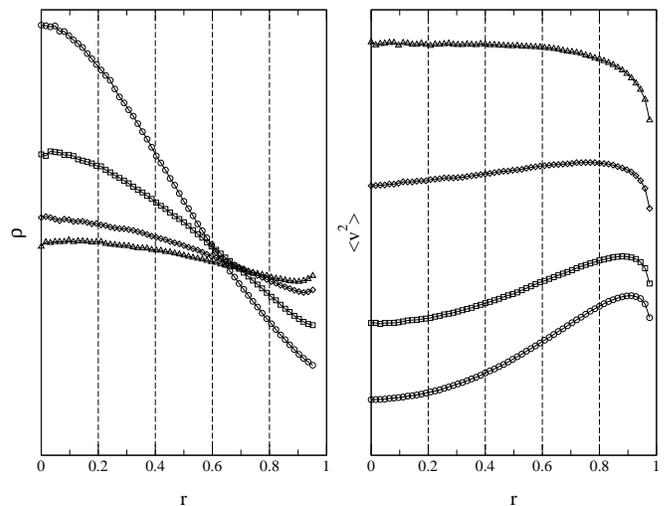} \caption{(a) The average number density $\rho$ as a function of distance $r$ to the center of the box. Data taken for $N=350$, $\eta=0.9$ and $\phi=0.1$($\circ$), $0.05$($\square$), $0.02$($\diamond$) and $0.01$($\triangle$). (b) The mean kinetic energy $\langle v^2\rangle$ per particle as a function of $r$, for the same values of $\phi$ and $\eta$. Note that even for the dilute case $\phi=0.01$ the mean kinetic energy profile is not constant, but drops at the boundary of the box. The profile only becomes constant after a certain distance into the container, that corresponds to the mean free path of particles leaving the boundary.} \label{fig:profile_eta} \end{figure}
\begin{figure}
    \centering
    \includegraphics[width=\columnwidth]{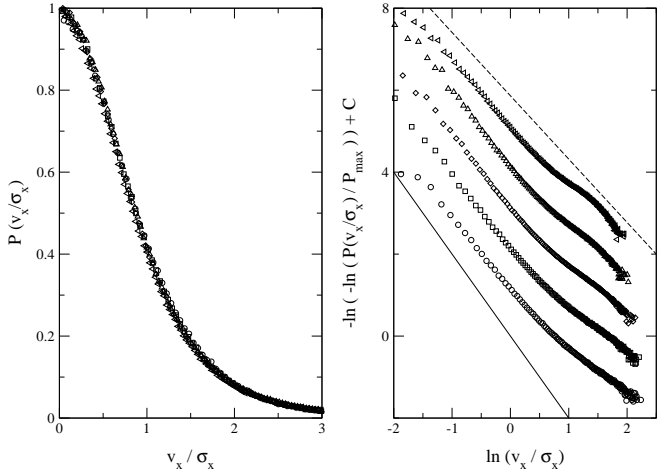}
\caption{Velocity distributions calculated separately within the concentric rings shown in Figs.\ \ref{fig:profile_phi} and \ref{fig:profile_eta}, {\em i.e.}, for $0<r\leq0.2$($\circ$), $0.2<r\leq0.4$($\square$), $0.4<r\leq0.6$($\diamond$), $0.6<r\leq0.8$($\triangle$) and $0.8<r\leq1$($\triangleleft$), where $r$ is distance to the center. Data were taken for $N=350$, $\phi=0.05$ and $\eta=0.8$. (a) $P(v/\sigma_x)$. (b) $-\ln\{-\ln[P(v_x/\sigma_x)]\}$ versus $\ln(v_x/\sigma_x)$. } \label{fig:boundaryexample}
\end{figure}

For boundary heating the gas develops a gradient in both density and mean kinetic energy as shown in figures \ref{fig:profile_phi} and \ref{fig:profile_eta}. Ideally, we want to measure velocity distributions in a region where the gradient is small. To this end we divided the box in five rings of width $0.2$. These rings are indicated in figures \ref{fig:profile_phi} and \ref{fig:profile_eta}. Only for values of $\phi$ and $\eta$ close to the clustering state, does the density within a ring vary by more than $10\%$. The velocity distributions $P(v_x)$ for particles within the different rings are shown in figure \ref{fig:boundaryexample}. Figure \ref{fig:boundaryexample}(a) shows $P(v_x)$ with the velocity component $v_x$ scaled by $\sigma_x=\langle v^2_x\rangle ^{\frac{1}{2}}$ and the maximum of the distribution scaled to be unity. All distributions seem to have the same shape. This is remarkable, since naively one would expect the dynamics of the particles to be different for the different rings. In the center the density is high and particles have many collisions, causing strong spatial correlations. Along the boundary the gas is dilute. Here, many particles have experienced recent collisions with the boundary, and spatial correlations should be much weaker. It is has been suggested that these correlations cause the velocity distributions to deviate from Gaussian. But contrary to the expectations, within the container the shape of the distributions seems to be determined by the granular temperature $\sigma_x$ only. This is a feature that is observed for all values of $\phi$ and $\eta$, even close to the cluster state.

\begin{figure}
    \centering
    \includegraphics[width=\columnwidth]{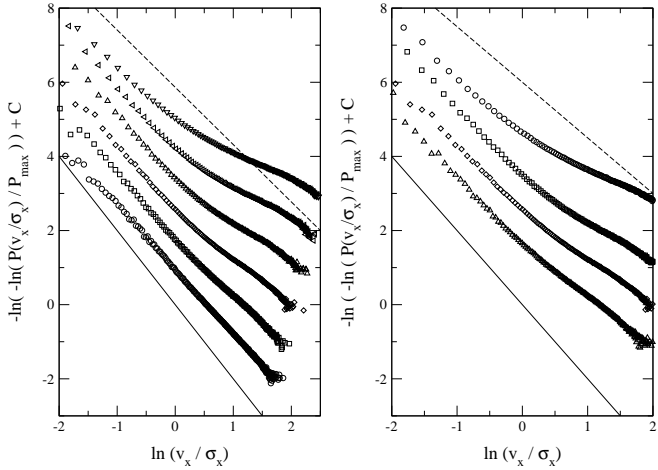}
\caption{(a) $-\ln\{-\ln[P(v_x/\sigma_x)]\}$ versus
$\ln(v_x/\sigma_x)$ for $N=350$, $\phi=0.02$ and
$\eta=0.9$($\circ$), $0.8$($\square$), $0.7$($\diamond$),
$0.6$($\triangle$), $0.5$($\triangleleft$) and
$0.4$($\bigtriangledown$). (b) $-\ln\{-\ln[P(v_x/\sigma_x)]\}$
versus $\ln(v_x/\sigma_x)$ for $N=350$, $\eta=0.7$ and
$\phi=0.01$($\circ$), $0.02$($\square$), $0.03$($\diamond$) and
$0.05$($\triangle$).} \label{fig:boundary2}
\end{figure}

Figure \ref{fig:boundaryexample}(b) shows the behavior of the exponent $\alpha$. This behavior is very different from the case of uniform heating. For uniform heating $\alpha$ has the same value over the entire observed range of velocities. For boundary heating, on the other hand, $\alpha$ has a constant value $\alpha_1$ over the low-velocity range but crosses over to different value $\alpha_2$ when above a critical velocity $v_c$. For the two innermost rings $\alpha_1=1.7$ and $\alpha_2=1.0$. For the third ring $\alpha_1=1.8$ and the outer ring has $\alpha=1.5$. For the outer two rings the velocity distributions actually undergo two crossovers. First it crosses over to $\alpha_2 \approx 1.0$ and then the exponent increases again to $\alpha_3 \approx 1.5$. In figure \ref{fig:boundary2} we show the effect of a change in $\phi$ and $\eta$ on the shape of the velocity distributions. Here we focus on the velocity distribution as measured in the ring with $0.4 <r\leq 0.6$. This has the advantage of good statistics, but for values of $\phi$ and $\eta$ close to a cluster, we might see effects due to the density gradient in the gas. As shown in figure \ref{fig:boundary2}(a) the exponent $\alpha_1=1.8$ except for $\eta=0.4$, where $\alpha_1=1.6$. For $\eta=0.9$ there is no crossover in the observed range of velocities. In the other distributions one does observe a crossover and the point where it occurs shifts down to lower velocities as $\eta$ is decreased. It is clear that the distribution for velocities above the crossover cannot be described by a single exponent. For low enough $\eta$, the distribution seems to approach a constant exponent for high velocities. This exponent decreases from $\alpha_2=1.3$ for $\eta=0.7$ to $\alpha_2=1.0$ for $\eta=0.4$. In figure \ref{fig:boundary2}(b) $\alpha_1=1.8$ for $\phi=0.01$ and $\phi=0.02$, $\alpha_1=1.9$ for $\phi=0.03$ and $\alpha_1=1.7$ for $\phi=0.05$. For every $\phi$ does one observe a crossover and the velocity at which the crossover occurs shifts only a bit as $\phi$ is varied. The distributions approach a constant exponent for high velocities. This exponent goes down from $\alpha_2=1.5$ for $\phi=0.01$ to $\alpha_2=1.0$ for $\phi=0.05$. In general, the deviations from Gaussian become more pronounced as dissipation increases, i.e. as $\phi$ increases or as $\eta$ decreases.

To test whether the velocity distributions we find here are only observed for this specific driving mechanism of heating through a circular boundary, we constructed different systems that drive through boundaries in a different way. For instance, we constructed a box with periodic boundary conditions that includes a small circular region around the center. Within this circular region particles are uniformly driven but outside of the region they are not heated at all. For particles within the circular region we observe velocity distributions that are Gaussian. On the other hand, for particles outside of the circular region we observe the same non-Gaussian velocity distributions as seen in the case of a circular boundary. 

For uniform heating the distribution of velocities of particles that are heated is Gaussian, because the velocities undergo a random walk. This is not the case for the particles heated at the boundary. To see if velocity distributions are sensitive to the precise way of heating at the boundary, we changed our heating algorithm so that, when a particle hits a boundary, it's new velocity is drawn from a Gaussian distribution. This has a modest effect on the far end of the high-velocity tail, but leaves all major differences between uniform and boundary heating intact. 

\begin{figure}
  \centering
  \includegraphics[width=\columnwidth]{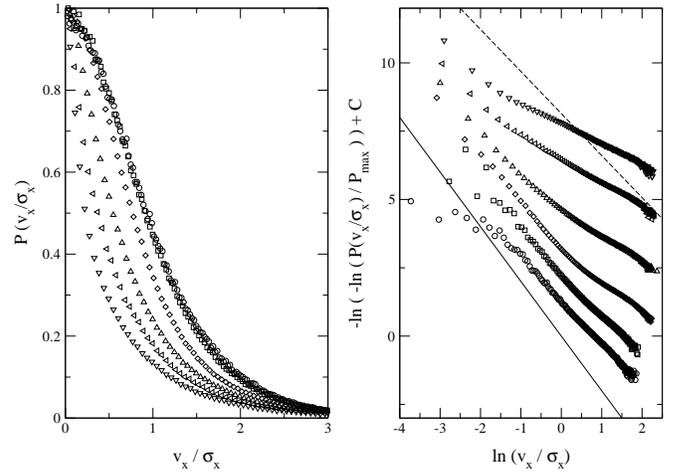}
\caption{(a) $P(v_x/\sigma_x)$ for different values of $N$. (b)
$-\ln\{-\ln[P(v_x/\sigma_x)]\}$ versus $\ln(v_x/\sigma_x)$. Data
is taken for $\phi=0.05$, $\eta=0.8$ and $N=100$($\circ$),
$200$($\square$), $500$($\diamond$), $700$($\triangle$) and
$1000$($\triangleleft$).} \label{fig:diffN}
\end{figure}

Finally, we studied the behavior of the velocity distribution for different particle numbers $N$. Figure \ref{fig:diffN}(a) shows that, as $N$ increases, the velocity distributions fall off more rapidly, but the high-velocity tails grow longer. This is shown more clearly in figure \ref{fig:diffN}(b). For the high-velocity tail we find $\alpha=1.7$ for $N=100$ decreasing to $\alpha=0.7$ for $N=1000$. The crossover shifts to lower velocity and becomes sharper as $N$ is increased. As we will demonstrate in more detail below, for boundary heating there is no thermodynamic limit. Instead of approaching a limiting velocity distribution as $N$ is increased, we find that the shape of the velocity distribution depends not only on $\eta$ and $\phi$, but also on $N$.

\begin{figure}[hb]
  \centering
  \vspace{7mm}
  \includegraphics[width=\columnwidth]{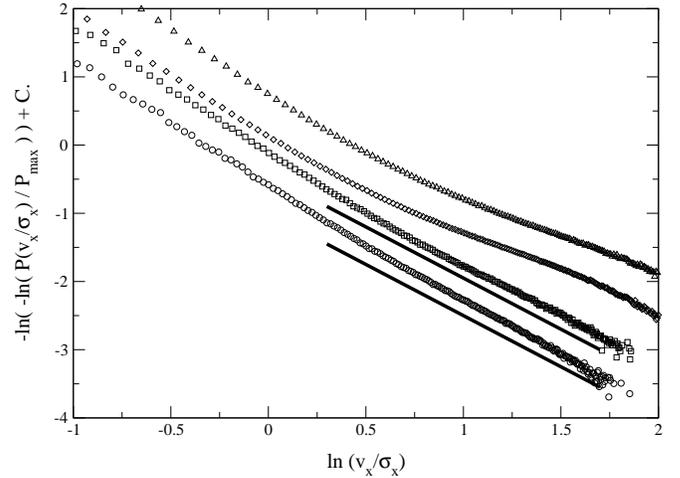}
\caption{(a) $-\ln\{-\ln[P(v_x/\sigma_x)]\}$ versus $\ln( v_x/\sigma_x)$ for $N=350$, $\phi=0.05$ and $\eta=0.9$ ($\circ$), $N=500$, $\phi=0.05$ and $\eta=0.9$ ($\square$), $N=350$, $\phi=0.05$ and $\eta=0.8$ ($\diamond$), $N=350$, $\phi=0.25$ and $\eta=0.9$ ($\triangle$). The solid lines correspond to the fit as made by Rouyer and Menon and has an exponent $\alpha=1.52$. The range of the solid lines corresponds to half the range used by Rouyer and Menon in their fit, but contains about 80$\%$ of their data points}. \label{fig:rouyer}
\end{figure}

For their experiments Rouyer and Menon used $N$ particles with $\eta\approx0.9$, where $100<N<500$ and $0.05<\phi<0.25$ \cite{rouyer}. In figure \ref{fig:rouyer} we plotted the velocity distribution for $\eta=0.9$, $\phi=0.05$ and several values of $N$. The solid black lines indicated the fit with $\alpha=1.52$ as made in \cite{rouyer}. This line clearly coincides with the high-velocity tail of the velocity distribution found by  simulation. This suggests that it is possible that instead of a universal distribution with $\alpha=1.5$, they only observed a part of a more complex velocity distribution, with two exponents and a crossover. For higher $\phi$ the  high-velocity tails have exponents that are smaller than the $\alpha=1.5$ observed for $\phi=0.05$.

\begin{figure}[hb]
  \centering
  \includegraphics[width=\columnwidth]{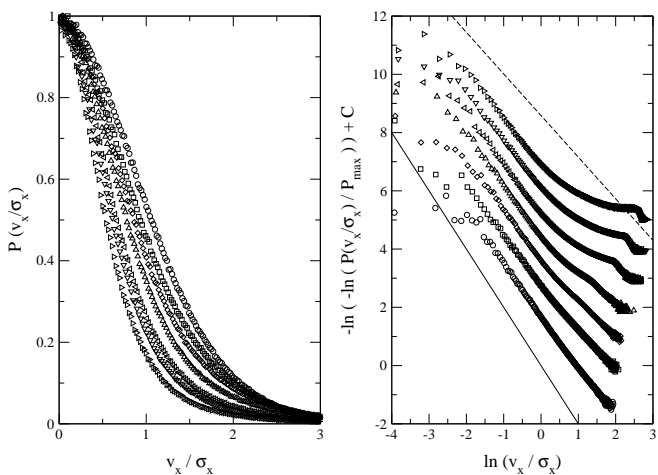}
\caption{(a) $P(v_x/\sigma_x)$ for different values of $\Delta t$. (b) $-\ln\{-\ln[P(v_x/\sigma_x)]\}$ versus $\ln( v_x/\sigma_x)$. Data is taken for $N=350$, $\phi=0.02$, $\eta=0.4$ and $\Delta t=0.01$($\circ$), $0.03$($\square$), $0.05$($\diamond$), $0.10$($\triangle$), $0.30$($\triangleleft$),
$0.50$($\triangledown$) and $1.00$($\triangleright$).}
\label{fig:2point}
\end{figure}

The main difference between uniform and boundary heating is that in the first case heating takes place homogeneously throughout the box, whereas in the latter case energy is injected inhomogeneously at the boundaries. This is not the direct cause for the difference in velocity distributions. This is demonstrated by the occurrence of non-Gaussian velocity distributions in a system studied by Moon et. al. \cite{moon}. They study a uniformly driven gas, but nevertheless find velocity distributions with a clear crossover. The main difference is the way they implement uniform heating. Every time $\Delta t$, they select at random two particles and add to these particles a random but opposite velocity to conserve the total momentum. On average heating is spatially homogeneous and in the limit of very small $\Delta t$ the behavior of both their and our uniform heating algorithm is the same. For small enough $\Delta t$ the number of collisions between heating events is smaller than one and the heating dominates over the dissipative behavior of the gas. When $\Delta t$ becomes larger, many collisions might happen between two heating events. Apperently this changes the velocity distributions. In figure \ref{fig:2point} we show the velocity distribution for $N=350$, $\phi=0.02$ and $\eta=0.4$. The gas is heated using the two-point heating algorithm described above, varying the time between heatings, $\Delta t$. For $\Delta t=0.01$ the distribution has a exponent $\alpha=1.7$ that is constant approximately over the observed range. When $\Delta t$ is reduced a clear crossover develops. The behavior of the velocity distribution for velocities higher than the crossover velocity is more complicated than in boundary heating. There is also a sharp kink at the high-velocity end, that we have been unable to explain so far.

This can explain the differences between velocity distributions for uniform and boundary heating. When heating uniformly all the particles are heated every timestep $\Delta t$, while only a few collisions occur in that time. The system is dominated by heating and we see a velocity distribution with a constant exponent over the entire observed velocity range. If we are heating through a boundary, there are many collisions in the dense region in the center of the box, but only a few particles escape to the boundary to get heated. In this case, the system is dominated by dissipation and we see in general a velocity distribution that strongly deviates from a Gaussian. This difference between uniform and boundary heating can be quantified by the ratio $q=N_H/N_C$, where $N_H$ and $N_C$ are the mean number of heatings and the number of collisions a particle has experienced, respectively. For a uniformly heated gas of $N=350$, $\phi=0.02$ and $\eta=0.4$ we find that $q=120$. For the same parameters but heating through the boundary we find $q=0.08$. For again the same parameters but using two-point heating with $\Delta t=1.0$ we find $q=0.012$.

\begin{figure}[ht] \centering 
\includegraphics[width=\columnwidth]{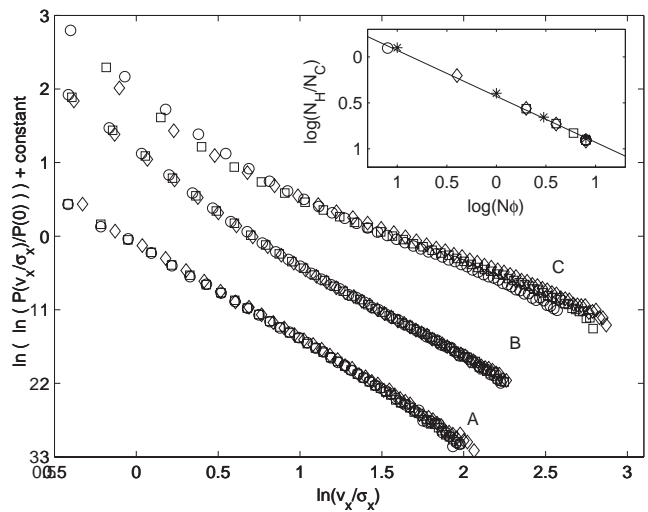} \caption{ Velocity distributions for different values of the heating-dissipation rate $q$. Distributions with the same $q$ are shown on top of each other. (A) $q=1.3$ and we show $N=100$ and $\phi=1\cdot10^{-3}$($\circ$), $N=200$ and $\phi=5\cdot10^{-4}$($\square$), $N=800$ and $\phi=1.25\cdot10^{-4}$($\diamondsuit$). (B) $q=0.13$ and we show $N=100$ and $\phi=0.08$($\circ$), $N=200$ and $\phi=0.04$($\square$), $N=400$ and $\phi=0.02$($\diamondsuit$). (C) $q=0.013$ and we show $N=100$ and $\phi=0.4$($\circ$), $N=200$ and $\phi=0.2$($\square$), $N=400$ and $\phi=0.1$($\diamondsuit$). Inset: Heating-dissipation ratio $q$ for $N=800$($\circ$), $N=400$($\square$), $N=200$($\diamondsuit$) and $N=100$($*$) for several values of $\phi$. The line is a fit of the form $(N \phi)^{1/2}$.}  \label{fig:collapse}
\end{figure}

We tested the effect of changing $q$ on velocity distributions obtained with boundary heating. When increasing the number of particles $N$ or the area fraction $\phi$, the average number of collisions increases. One can show in a mean field approximation that $q \sim (N \phi)^{-1/2}$. The average distance a particle travels between collisions is given by $l_{coll} \sim 1/\phi$. The average distance a particles travels between heatings $l_{heat}$ is given by the dimensions of the container. For a box of area $A$ the average distance between boundaries is given by $l_{heat} \sim A^{1/2}=(N/\phi)^{1/2}$. Finally, we know $N_H/N_C \sim l_{coll}/l_{heat}$. Our simulation obeys this approximation very well. In Fig.~\ref{fig:collapse} we show velocity distributions for $\eta=0.8$ and different combinations of $N$ and $\phi$. We measure the heating-dissipation ratio $q$ in the simulation and show velocity distributions with the same $q$ on top of each other. For $q=1.3$ and $q=0.13$ we find excellent collapse for different $N$ and $\phi$, even when we scale the system by a factor $8$. For $q=0.013$, where spatial correlations become very strong, we still find reasonable collapse. As we increase $q$ we observe the usual pattern, where a crossover appears in a distribution that was initially close to a Gaussian. This means that the velocity distributions are almost exclusively controlled by two parameters; the coefficient of restitution $\eta$, which is a material parameters, and the heating-dissipation ratio $q$, which depends on experimental conditions.

\section{a simple model without spatial degrees of freedom}

To study the dependence of the velocity distributions on the ratio $q$ we constructed a model of the inelastic gas inspired by work done by Ulam \cite{ulam} on elastic gases. Ulam rederived the Maxwell-Bolzmann distribution for the velocities of particles in a perfect gas using the following very simple model: he considered $N$ particles with random initial velocities. At every time step a pair of particles was selected and the velocities of the particles were changed as if they had collided elastically. Ulam found that, regardless of the initial velocities, the system quickly evolved to a state where on average all energy was distributed equally over all particles and the deviation of the velocities of the particles around the average velocity of the system was given by the Maxwell-Bolzmann distribution. This approximation of random collisions is only justified when the gas is sufficiently dilute. The same approximation can also be made for the granular gas by replacing the elastic collisions between particles with inelastic collisions. Since the gas cools down without any further energy input, heating has to be included as well. This allows us to study the dissipative behavior of the granular gas while neglecting all spatial correlations that develop in the more detailed simulations described before. In particularly we can study the effect on the velocity distributions of changing the fraction $q=N_H/N_C$ described in the previous section by varying the ratio of heatings and collisions.

We adapted Ulam's procedure in order to model a two-dimensional granular gas consisting of $N$ particles. Equation \ref{eqn:inelas} can be cast into the following form:
\begin{equation}
\mathbf{v}_i^{\prime}=\mathbf{v}_i-\frac{1+\eta}{2}
\left( \begin{array}{cc}
\cos^2 \theta & \sin \theta \cos \theta \\
\sin \theta \cos \theta & \sin^2 \theta
\end{array} \right)
(\mathbf{v}_i-\mathbf{v}_j),
\end{equation}
where $\mathbf{v}_i$ and $\mathbf{v}_i$ are the velocities of particles $i$ and $j$, $\eta$ is the coefficient of restitution and $\theta$ is the angle between the separation vector $\mathbf{r}_{ij}$ and a reference axis. Collisions in our model occur by selecting at random particles $i$ and $j$ and an uniformly distributed impact parameter $-2R<b<2R$. We then use the above collision rule with
$\theta=\arcsin(b/2R)+\arccos(\mathbf{v} \cdot \hat{\mathbf{s}}/v)$, where $\mathbf{v}=(\mathbf{v}_j-\mathbf{v}_i)/2$ is the velocity in the center-of-mass frame and $\hat{\mathbf{s}}$ is a unit vector along the reference axis. We discard values of $\theta$ corresponding to $(\mathbf{v}_j-\mathbf{v}_i) \cdot \mathbf{r}_{ij} < 0$ as these represent unphysical collisions. We heat the gas by selecting a random particle $i$ and adding a random amount of velocity according to equation \ref{eqn:uniheat}. To prevent the velocities from running away, we subtract the center-of-mass velocity after heating. In a single time step, we have let $2C$ particles collide and we heat $H$ particles. This gives us $q=H/(2C)$.

When we use inelastic instead of elastic collisions and drive the system, we find that the gas heats up until it reaches a steady state, where on average the energy lost in collisions is compensated by the heat inserted into the system through our driving mechanism. As $H$ is chosen increasingly bigger in comparison to $C$, the granular temperature $\sigma$ of the gas increases accordingly.

\begin{figure}[hb]
\begin{center}
\vspace{5mm}
\includegraphics[width=\columnwidth]{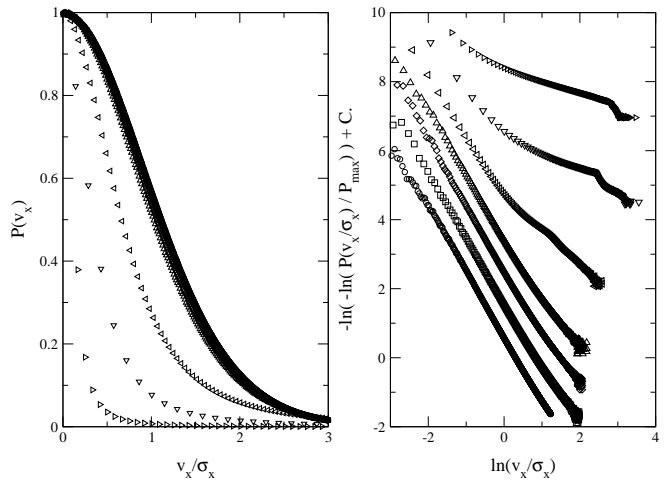} \caption{
Velocity distribution for an inelastic gas of $N=500$ and $\eta=0.4$. (a) $P(v_x/\sigma_x)$. (b) $-\ln\{-\ln[P(v_x/\sigma_x)]\}$ versus $\ln(v_x/\sigma_x)$. Data is for different values of $q=\frac{H}{2C}$: $q=50$ ($\circ$), $q=5$ ($\square$), $q=1$ ($\diamond$), $q=0.5$ ($\triangle$), $q=0.05$ ($\triangleleft$), $q=5\cdot 10^{-3}$ ($\triangledown$) and $q=5\cdot 10^{-4}$ ($\triangleright$).} \label{fig:diff_f} 
\end{center}
\end{figure}

\begin{figure}[hb]
\begin{center}
\vspace{0.3cm} \includegraphics[width=\columnwidth]{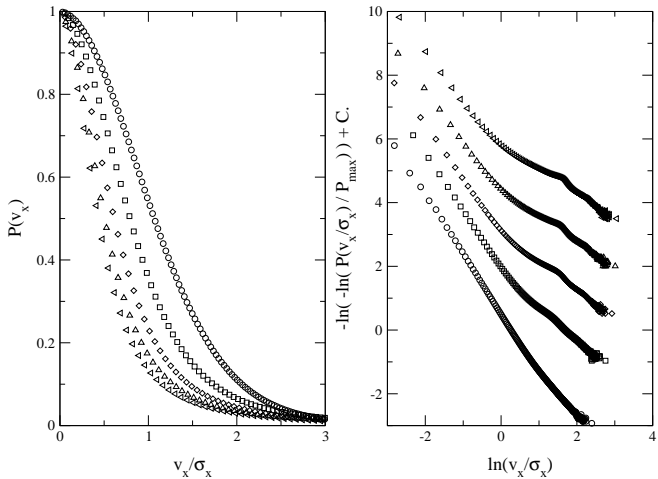}
\caption{ Velocity distribution for an inelastic gas of $N=500$ and $q=0.025$. (a) $P(v_x/\sigma_x)$. (b) $-\ln\{-\ln[P(v_x/\sigma_x)]\}$ versus $\ln(v_x/\sigma_x)$. Data is for $\eta=0.9$ ($\circ$), $0.7$ ($\square$), $0.5$ ($\diamond$), $0.3$ ($\triangle$) and $0.1$ ($\triangleleft$).}
\label{fig:diffeta}
\end{center}
\end{figure}

In figure \ref{fig:diff_f} we plotted the result for an inelastic gas with $\eta=0.4$. We varied the number of heatings and the number of collisions in a single time step from $H=100$ and $C=1$ to $H=1$ and $C=1000$. As $q$ is lowered, the velocity distributions develop a crossover and for $q \ll 1$ the distributions are strongly non-Gaussian, similar to the velocity distributions obtained for $\Delta t \gg 1$ in two-point heating. In figure \ref{fig:diffeta} we keep $q=0.025$ fixed and vary $\eta$. We see that the crossover point shifts down as $\eta$ is lowered and that the kink in the velocity distribution shifts down.

\begin{figure}[hb]
\begin{center}
\vspace{.5cm}
\includegraphics[width=\columnwidth]{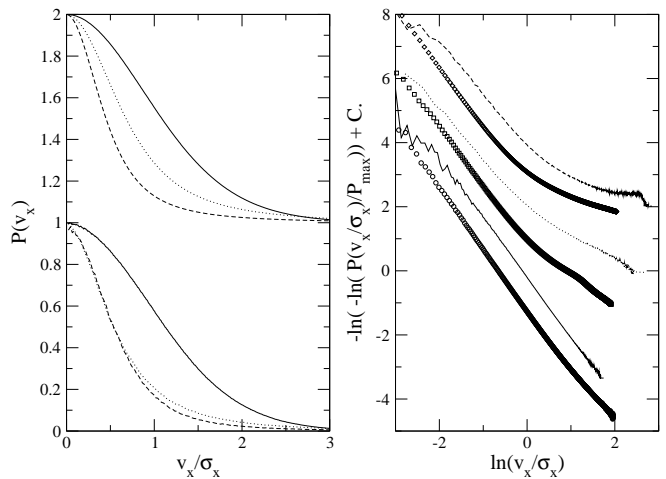}
\caption{ Velocity distributions for an inelastic gas of $N=350$, $\phi=0.02$ and $\eta=0.4$. (a) $P(v_x/\sigma_x)$ for $q=120$ (solid), $q=0.08$ (dashed) and $q=0.012$ (dotted). Results on the bottom are obtained for simulation, results on the top for the model. (b) $-\ln\{-\ln[P(v_x/\sigma_x)]\}$ versus $\ln(v_x/\sigma_x)$. The symbols shown are velocity distributions acquired by simulation for $q=120$ ($\circ$), $0.08$ ($\square$), $0.012$ ($\diamond$). The lines show the velocity distributions found in the model for the same values of $q$ (solid, dotted, dashed).} \label{fig:fcompare}
\end{center}
\end{figure}

We also compared velocity distributions found in the model with those acquired by simulation. To this end we measured $q$ in simulations using different ways of heating. For a uniformly heated gas of $N=350$, $\phi=0.02$ and $\eta=0.4$ we found $q=120$. For the same parameters but heating through the boundary we found $q=0.08$. For two-point heating with $\Delta t=1.0$ we found $q=0.012$. We ran the model with the corresponding set of parameters and measured velocity distributions. The results are shown in figure \ref{fig:fcompare}.

The velocity distributions agree qualitatively, but especially for the case of q=0.08 there is considerable quantitative difference. Still, the model is able to generate non-Gaussian  velocity distributions, while neglecting all spatial correlations. This suggests that the non-Gaussian behavior of the velocity distributions is not caused by spatial fluctuations and correlations in the gas. Instead it is the flow of energy through the system, mediated by the inelastic collisions, that determines the shape of the velocity distribution. This also means that there is no a priori difference between inhomogeneous and homogeneous heating: the difference in the shape of the velocity distribution, particularly the occurrence of the crossover, is only a consequence of the fact that in inhomogeneous heating in general the number of collisions between particles exceeds the number of particles being heated.

\section{relation to experiments}

Due to the idealized nature of our system, it is not possible to do a direct comparison between our simulation and experiments. In the experiment of Rouyer and Menon the driving is through the boundary, but there are some significant differences between their heating mechanism and the one we use in simulations with boundary heating. Due to gravity and the geometry of the setup the injection of energy in the experiment of Rouyer and Menon is mainly in the vertical direction. This energy is transferred into the horizontal direction by collisions between particles. Another difference is that in the experiment the frequency of driving is relatively low. Because of this the dynamics of the gas close to the driving boundary is strongly dependend on the phase of the driving cycle. In fact, it has been shown in simulation that for a system similar to the experiment by Rouyer and Menon, a shockwave propagates up through the gas \cite{bougie}. At a certain distance from the boundary the time dependence has decayed away and the gas enters a steady state. It is in this steady state that the velocity distributions are measured. 

It is not yet established how this time dependence and the occurence of a shock wave influences the velocity distributions in the steady state. A priori it is not clear if it is possible to compare velocity distributions in systems with a strong time-dependence, like the experiments, with those that have no time dependence, as is the case in our simulations. There are, however, reasons to assume this is possible. The velocity distributions in the experiment of Rouyer and Menon are measured only in the direction orthogonal to the driving direction. Simulations \cite{bougie} show that the effect of the shock in the direction orthogonal to the shock are usually relatively weak and decay rapidly in height. In the steady state, influence of the shock is absent in the orthogonal direction, even while it still may be apparent in the direction perpendicular to the driving direction. So, if we only look at velocity distributions in the orthogonal direction and in the steady state, a comparison between the experiment and our simulations is be valid.

It is also not clear how in the experiments the dynamics of the gas shape the velocity distribution and whether it is controlled by the parameter $q$. We speculate that in the steady state the system behaves in fact like a one dimensional inelastic gas. Fast upward moving particles inject energy in the orthogonal direction when colliding with particles in the steady state, effectively functioning as a heat source. In this picture the mean number of collisions between fast upwards moving particles and particles in the steady state would be $N_H$, the average number of heatings, and collisions between the particles in the steady state mutually would be $N_C$, the average number of collisions. One way of changing the shape of the velocity distribution would be changing the fraction of particles in the steady state. More particles in the steady state would lower $N_H$ and increase $N_C$ leading to more non-Gaussian velocity distributions.

The above considerations not only apply to the experiment of Rouyer and Menon but to most of the other experiments as well. In the experiments of Blair and Kudrolli and those of Olafsen and Urbach velocity distributions are measured orthogonal to the driving direction. In both cases it is not so much the collisions with the bottom plate that drive the gas in the ortogonal directions, but mainly collisions between fast upward moving particles with particles that have low velocities in the orthogonal directions. It is in these experiments rather than those of Rouyer and Menon that we find a similar dependence on $\eta$ and $q$ as we describe in this article. 

In the setup of Olafsen and Urbach \cite{olafsen} velocity distributions go from non-Gaussian to Gaussian when a rough plate is used instead of a flat plate. On a flat plate, energy is only injected in the in-plane directions by off-angle collisions between neighboring particles. With a rough plate, energy is injected directly into the directions parallel to the plate every time a particle collides with the plate, effectively increasing the number of heatings over collisions. Baxter and Olafsen \cite{baxter} observe the same behavior in a system where a layer of heavy particles is inserted between the other layer of particles and a flat bottom plate. Particles from the upper layer have off-angle collisions with the layer of heavy particles, injecting energy in the in-plane directions every cycle. Particles in the upper layer show Gaussian velocity distributions, whereas particles in the lower layer have non-Gaussian velocity distributions.
 
Most convincing is the experiment by Blair and Kudrollli \cite{kudrolli}. Here the number of collisions is increased by adding more particles. As a result, their velocity distributions develop the same crossover we see both in our simulations and model. The reason why these transitions are not visible in the experiment by Rouyer and Menon as they increase number of particles, is that in the first case the effective coefficient of restitution is much lower, $\eta \approx 0.5$, due to friction with the inclined plane.   

Again, one of the main problems considering the velocity distributions in granular gases is that different setups and experiments usually find different velocity distributions. As we have shown in this section, our finding of the controlling parameter $q$ could ultimately explain these seemingly inconsistent results.

\section{conclusion}

We compared the velocity distributions of a granular gas that was driven by uniform heating and by heating through the boundary. Although it is usually assumed that uniform heating yields the same behavior as boundary heating when the boundary is far away and the spatial gradients are small, we find that there are clear qualitative differences. When driven through the boundary, for instance, the gas can form coexisting cool liquid-like clusters surrounded by a hot gaseous state for certain values of $\phi$ and $\eta$. Such clusters do not occur in our simulations with uniform heating. 

The difference between uniform heating and boundary heating also extends to the velocity distributions. For uniform heating, we find velocity distributions that are close to Gaussian over the observed range of velocities and for a wide range of $\phi$ and $\eta$. When heating through the boundaries, on the other hand, we find that velocity distributions are stretched Gaussians that usually cross over from one exponent $\alpha$ to another for the high-velocity tail. For the high-velocity tail we find that the exponent $\alpha$ varies over a wide range. For strong dissipation this tail cannot be described by a single exponent.

This means that there is no universal velocity distribution with $\alpha=1.5$, as was proposed by Rouyer and Menon. Instead, the velocity distribution found by Rouyer and Menon may be just one out of the many distributions described here. The apparent universality may be due to the use of a narrow range of parameters (nearly elastic particles were used with $\eta=0.93$ and vary $0.05\leq\phi\leq 0.25$). We find that varying $\phi$ only has a minor effect on the shape of the distribution when $\eta$ is so close to the elastic limit. For low $\eta$, we instead find velocity distributions that look much like the distributions found by Blair and Kudrolli for gases with $\eta \approx 0.5$. 

Furthermore, we demonstrate that the distribution of velocities for dissipative gases, while not universal in form, depends only on two parameters: the coefficient of restitution $\eta$ (a material parameter) and $q=N_H/N_C$, the average ratio of heatings and collisions in the gas (a function of experimental conditions). We find that velocity distributions range from Gaussian for $q \gg 1$, where heating dominates dissipation, to strongly non-Gaussian for $q \ll 1$, where the dynamics of the gas is dominated by the dissipative collisions between particles. The differences in distributions we find between uniform and boundary heating can then be understood as different limits of $q$, for $q \gg 1$ and $q \lesssim 1$ respectively. We can control the parameter $q$ more easily using two-point heating and here we observe the transistion form Gaussian to strongly non-Gaussian behavior directly as we change $q$.

Furthermore, a simple model of a driven, inelastic gas without spatial degrees of freedom reproduces the entire family of velocity distributions we find in simulation, as we vary $\eta$ and $q$. This means that the velocity distributions are non-Gaussian not because of spatial correlations. Rather, it is the cascade of energy from a few high-energy particles to the slow-moving bulk of the gas that is the key determinant of the non-Gaussian velocity distributions. These observations should aid in the construction of a kinetic theory of dissipative gases.

We thank L. P. Kadanoff, N. Menon, A. Kudrolli, D. Blair and J. Bougie for useful conversations. This work is supported in part by the National Science Foundation under Grant PHY99-07949.

\end{document}